\documentclass[aps,prl,showpacs,amsmath,amssymb,amsfonts,lengthcheck,twocolumn,longbibliography]{revtex4-1}

\usepackage{graphicx}
\usepackage{subfigure}
\usepackage{verbatim}
\usepackage{dcolumn}
\usepackage{bm}
\usepackage{epsf}
\usepackage{color}
\usepackage[colorlinks=true,citecolor=blue,linkcolor=blue,urlcolor=blue]{hyperref}
\usepackage{hhline}

\newcommand{\tr}[1]{\mathrm{tr}\left\{#1\right\}}

\newcommand{\la}{\left\langle}
\newcommand{\ra}{\right\rangle}
\newcommand{\pd}{\partial}

\newcommand{\td}{\mathrm{d}}

\newcommand{\e}[1]{\exp{\left(#1\right)}}
\newcommand{\lo}[1]{\ln{\left(#1\right)}}

\newcommand{\bla}{bla\\bla\\bla\\bla\\bla}

\newcommand{\mb}[1]{\mbox{\boldmath$#1$}}
\newcommand{\mc}[1]{\mathcal{#1}}

\newcommand{\mrm}[1]{\mathrm{#1}}

\begin{document}

\title{Kibble-Zurek scaling of the irreversible entropy production}

\author{Sebastian Deffner}
\affiliation{Department of Physics, University of Maryland Baltimore County, Baltimore, MD 21250, USA}
\date{\today}

\begin{abstract}
If a system is driven at finite-rate through a phase transition by varying an intensive parameter, the order parameter shatters into finite domains. The Kibble-Zurek mechanism predicts the typical size of these domains, which are governed only by the rate of driving and the spatial and dynamical critical exponents. We show that also the irreversible entropy production fulfills a universal behavior, which however is determined by an additional critical exponent corresponding to the intensive control parameter. Our universal prediction is numerically tested in two systems exhibiting noise-induced phase transitions.
\end{abstract}


\maketitle

\paragraph*{Introduction}
If the Universe started with a Big Bang during which all mass and energy was concentrated in an infinitely small volume, how come that nowadays matter is so sparsely distributed? Realizing that the early Universe must have undergone a phase transtion, Kibble noted that relativistic causality alone makes the creation of topological defects and the existence of finite domain sizes inevitable \cite{Kibble1976}. In laboratory phase transitions, however, relativistic causality does not lead to useful insights \cite{Zurek1985}. 

In thermodynamics  second order phase transitions can be classified into \emph{universality classes} \cite{Callen1985a}. At the critical point thermodynamic response functions, such as the magnetic susceptibility, diverge, $\chi\sim |T-T_c|^{-\gamma}$, where $T$ is the temperature and $\gamma$ is called critical exponent. Typically,  $\gamma$ only depends on symmetries and not on microscopic details, and thus the values of $\gamma$ are \emph{universal} for classes of systems \cite{Fisher1974}.

The divergence of response functions at the critical point can be understood as a ``freezing out'' of all dynamics. It is exactly this critical slowing down in the vicinity of the critical point that allows to predict the density of defects, the size of typical domains, and their excitations \cite{Zurek1985,Zurek1996,Laguna1997}. The \emph{Kibble-Zurek mechanism} (KZM) has been very successfully tested in thermodynamic phase transitions \cite{Ruutu1996,Bauerle1996,Monaco2006}. More recently, a variety of experimental studies was reported, which were able to confirm the predictions of the phenomenological theory also, for instance, in trapped ions \cite{Ulm2013,Partner2013} and in Bose-Einstein condensates \cite{Scherer2007,Weiler2008}. Moreover, the KZM also has been extended to inhomogeneous systems \cite{delCampo2013c,DelCampo2014}, quantum phase transitions \cite{Zurek2005a,Damski2005,Dziarmaga2005,Damski2007,Dziarmaga2012}, and biochemical networks \cite{Erez2017}.

From a thermodynamic point of view, however, the picture appears to be not entirely complete. At the very core of the KZM is the understanding that driving a system at finite rate through a phase transition makes the system break up into finite domains. In particular, the faster a system is driven, the smaller are the pieces into which the order parameter is shattered \cite{Laguna1997}. The breaking of the system into finite-sized domains, however, has to be accompanied by dissipated work, or rather irreversible entropy production, $\la \Sigma\ra$. The natural question arises whether arguments of the KZM allow to determine $\la \Sigma\ra$, and whether also $\la \Sigma\ra$ exhibits universal behavior.

In the present analysis, we derive a general expression for $\la \Sigma\ra$ and show that it does, indeed, obey a universal scaling law. To this end, we combine concepts from \emph{Stochastic Thermodynamics} \cite{Seifert2012} , \emph{Finite-Time Thermodynamic} \cite{Salamon1983,Andresen1984}, and the KZM. More specifcially, we use that the irreversible entropy production can be written as quadratic form of the susceptibility $\chi$ \cite{Sivak2012a,Bonanca2014a} to show that
\begin{equation}
\label{eq01}
\la \Sigma\ra \sim \tau_Q^{\frac{\Lambda-2}{1+z\nu}}\,,
\end{equation}
where $\tau_Q$ is the quench time, $\nu$ the spatial, $z$ the dynamic critical exponent, and $\Lambda$ is the exponent corresponding to the varied parameter. If the system is driven by varying the magnetization we have $\Lambda=\gamma$, whereas for time-dependent temperatures we find $\Lambda=\alpha$.

Our general findings are illustrated with two elucidating systems. We corroborate the conceptual arguments with a numerical study of \emph{noise-induced phase transitions} \cite{Broeck1994,Toral2011} in mean-field description, for which notions such as ``domain sizes'' or ``number of defects'' are somewhat loose. Thus the present analysis not only closes the conceptual gap between the KZM and Stochastic Thermodynamics, but also significantly extends the applicability of the KZM to phase transitions between nonequilibrium states.

\paragraph*{Preliminaries: the Kibble-Zurek mechanism}

We begin by briefly reviewing the main notions of the KZM and establish notations. Close to the critical point both the correlation length, $\xi$, as well as the correlation time, $\tau$, diverge. Renormalization group theory predicts \cite{Fisher1974,Herbut2007a} that
\begin{equation}
\label{eq02}
\xi(\epsilon) = \xi_0\,\left|\epsilon\right|^{-\nu} \quad\mathrm{and}\quad\tau(\epsilon)=\tau_0\,\left|\epsilon\right|^{-z\nu}\,,
\end{equation}
where $\epsilon$ is a dimensionless parameter measuring the distance from the critical point, $\nu$ is the spatial and $z$ the dynamical critical exponent. In thermodynamic phase transtions $\epsilon$ is the relative temperature \cite{Zurek1985}, whereas in quantum phase transitions $\epsilon$ is a relative external field \cite{Zurek2005a,Francuz2015a}.

For the sake of simplicity we will assume that the system is driven through its phase transition by a linear ``quench''
\begin{equation}
\label{eq03}
\epsilon(t)=t/\tau_Q\,,
\end{equation}
and thus the constant quench rate $\dot{\epsilon}(t)$ is given by one over the quench time $\tau_Q$. Generalizing the KZM to nonlinear driving is straight forward and can be found, for instance, in Refs.~ \cite{Sen2008,Barankov2008,Mondal2009,Chandran2012,Chandran2013}.

\begin{figure}
\includegraphics[width=.48\textwidth]{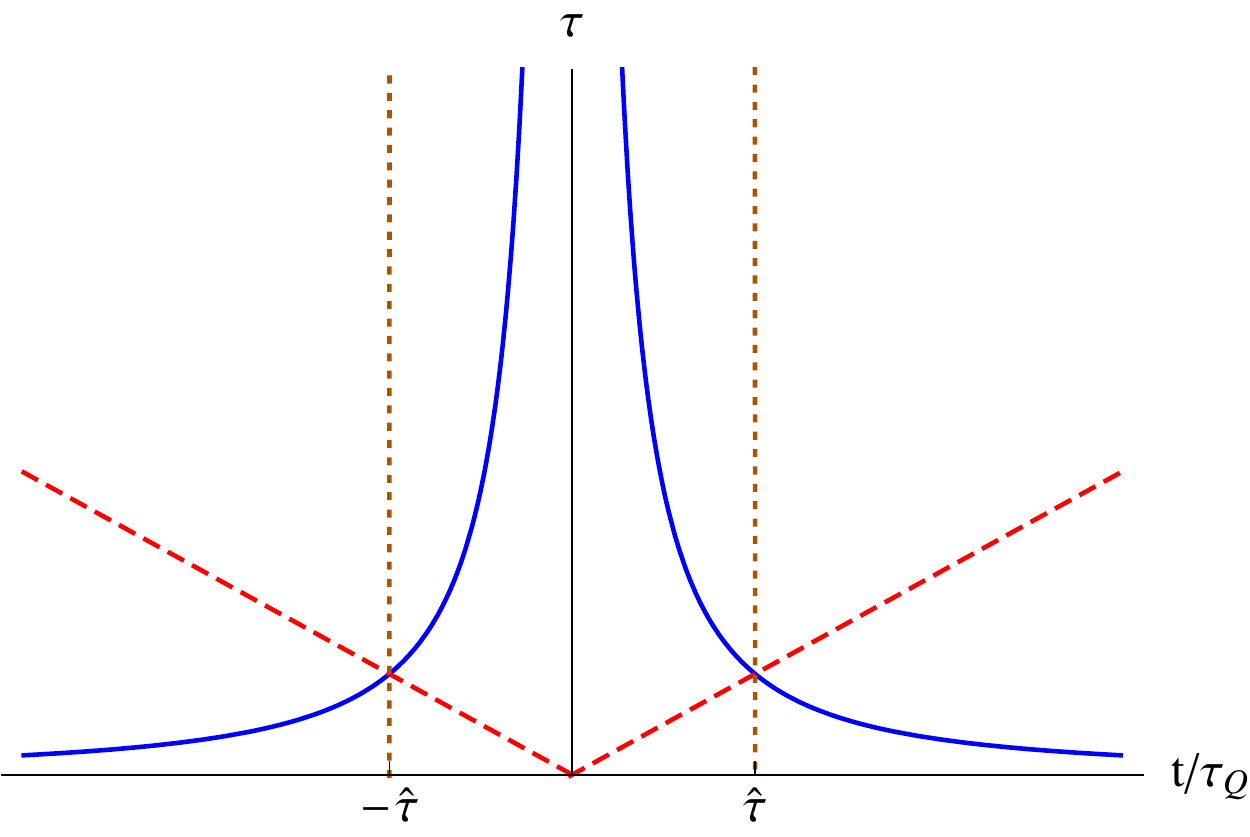}
\caption{\label{fig:plot_KZ}(color online) Relaxation time $\tau(t)$ \eqref{eq02} (blue, solid line) and rate of driving $|\dot{\epsilon}/\epsilon|$ (red, dashed line) for $\nu=1$ and $z=3/2$. The vertical lines illustrate the separation of the thermodynamic behavior into adiabatic and impulse regimes \cite{Zurek1996}.}
\end{figure}

For slow-enough driving and far from the critical point, $\tau\ll t$, the dynamics of the system is essentially adiabatic. This means, in particular, that all nonequilibrium excitations and defects equilibrate much faster than they are created. Close to the critical point, $\tau\simeq t$ the situation dramatically changes, since the response freezes out and defects and excitations cannot ``heal'' any longer.  This change of thermodynamic behavior, from adiabatic to ``impulse'' \cite{Zurek1996},  happens when the rate of driving becomes equal to the rate of relaxation, or more formally at
\begin{equation}
\label{eq04}
\hat{\tau}(\hat{t})=\hat{t}\quad\mrm{with}\quad\hat{\tau}=\left(\tau_0\,\tau_Q^{z \nu}\right)^\frac{1}{z\nu+1}\,.
\end{equation}
This insight is illustrated in Fig.~\ref{fig:plot_KZ}.

Accordingly the typical domain size is determined by the correlation length at $\hat{t}$, which can be written as,
\begin{equation}
\label{eq05}
\hat{\xi}=\xi(\hat{t})=\xi_0\,\left(\tau_Q/\tau_0\right)^\frac{\nu}{z\nu+1}\,.
\end{equation}
In many situations it is useful to introduce the density of defects $\rho_d$, which is given by the ratio $\hat{\xi}^d/\hat{\xi}^D$. Here $d$ and $D$ are the dimensions of defects and the space they live in, respectively. Thus, we can write,
\begin{equation}
\label{eq06}
\rho_d=\hat{\xi}^{(d-D)}\sim\tau_Q^{-\frac{(d-D)\,\nu}{z\nu+1}}\,,
\end{equation}
which sometimes is also called \emph{KZ-scaling}. It is important to emphasize that Eq.~\eqref{eq06} quantifies an effect of finite-rate, \emph{nonequilbirum} driving entirely in terms of the \emph{equilibrium} critical exponents. Note that in the original formulation of the KZM topological defects were considered since they constitute  robust signatures of the quench that can be easily counted. If, however, even correlation functions are accessible the scaling of the correlations length \eqref{eq05} can be directly measured.

The one question that is not addressed in this argumentation is whether the irreversible entropy production, $\la\Sigma\ra$, exhibits a similar behavior. Naively one would expect that per excitation the system is accompanied by a characteristic amount of entropy, $\sigma$,
\begin{equation}
\label{eq07}
\la \Sigma\ra \sim \rho_d\,\cdot\,\sigma\sim \tau_Q^{-\frac{d\,\nu}{z\nu+1}}\,.
\end{equation}
We will show in the following that this naive expectation is not entirely correct. Rather we will find that the behavior of the irreversible entropy production depends also on the critical exponent associated with the externally driven, intensive parameter.

\paragraph*{Maximum available work theorem}

The only processes that can be fully described by means of conventional thermodynamics are infinitely slow, equilibrium, aka \emph{quasistatic} processes \cite{Callen1985a}. Nonequilibrium processes are characterized by the \emph{maximum available work theorem} \cite{Schlogl1989}. Consider a general thermodynamic system which supplies work to a work reservoir, and which is in contact, but \emph{not in equilibrium} with a heat reservoir, $\mc{B}$. Then the first law of thermodynamics can be written as,
\begin{equation}
\label{app:eq01}
\Delta E+\Delta E_\mc{B}=\la W\ra\,,
\end{equation}
where $\Delta E$ is the change of internal energy of the system, $\Delta E_\mc{B}$ is the energy exchanged with $\mc{B}$, and as before $\la W\ra$ denotes the average work. Accordingly the second law of thermodynamics states \cite{Deffner2013},
\begin{equation}
\label{app:eq02}
\Delta S+\Delta S_\mc{B}\geq 0\,,
\end{equation}
where $\Delta S$ is the change of thermodynamic entropy of the system, $\Delta S_\mc{B}$ is the change of entropy in $\mc{B}$, and where we used that the entropy of the work reservoir is negligible \cite{Callen1985a,Deffner2013}. Since the heat reservoir is so large that it is always in equilibrium at inverse temperature $\beta$ we immediately can write $\beta \Delta E_\mc{B}=\Delta S_\mc{B}$, and hence we always have
\begin{equation}
\label{app:eq03}
\la W\ra\geq \Delta E-\Delta S/\beta\equiv \Delta \mc{E}\,.
\end{equation}
The thermodynamic quantity $\mc{E}$ is called exergy or availability \cite{Schlogl1989}, since it quantifies the maximally available work in any thermodynamic process.

\paragraph*{KZ-scaling of the excess work -- equilibrium systems}

The \emph{maximal available work theorem} \cite{Schlogl1989} can be re-written in terms of the excess work, $\la W_\mrm{ex}\ra$, which is given by the total work, $\la W\ra$, minus the quasistatic contribution, i.e., the availability $\Delta\mc{E}$, 
\begin{equation}
\label{eq08}
\la W_\mrm{ex}\ra=\la W\ra-\Delta \mc{E}\,.
\end{equation}
At constant temperature we can write $\Delta \mc{E}\equiv\la W\ra-\Delta E+\Delta S/\beta$ \cite{Schlogl1989}, where $\Delta E$ is the change of internal energy in a quastistatic process, $\Delta S$ denotes the change of entropy, and $\beta$ is the inverse temperature. For open equilibrium systems and isothermal processes the availability further reduces to the difference in Helmholtz free energy, $\Delta\mc{E}=\Delta F$, why we can also write $\la \Sigma\ra=\beta \la W_\mrm{ex}\ra$ \cite{Deffner2011b}. 

However, more generally $\Delta\mc{E}$ is the work performed during any quasistatic process, and thus $\la W_\mrm{ex}\ra$ quantifies the nonequilibrium excitations arising from finite time driving -- in isothermal as well as in more general processes, and in open as well as in isolated systems \cite{Allahverdyan2005a,Allahverdyan2007,Acconcia2014,Acconcia2015a,Bonanca2015}. 

Motivated by insights from finite-time thermodynamics \cite{Salamon1983,Andresen1984} it has recently become clear  that for sufficiently slow processes $\la W_\mrm{ex}\ra$ can be  expressed as quadratic form  \cite{Sivak2012a,Bonanca2014a},
\begin{equation}
\label{eq09}
\la W_\mrm{ex}\ra=\int dt\,\frac{d \mb\lambda^\dagger}{dt}\,\tau(t)\,\mc{I}(t)\,\frac{d \mb\lambda}{dt}\,,
\end{equation}
where $\mb{\lambda}=(T,V,H,\dots)$ is the vector of all intensive parameters varied during the process, such as temperature $T$, volume $V$, magnetic field $H$, etc., and the integral is taken over the whole process. Furthermore, $\mc{I}(t)$ is the Fisher information matrix, which  for a $d$ dimensional system close to the critical point and for only two intensive parameters such as $T$ and $H$ can be written as \cite{Prokopenko2011},
\begin{equation}
\label{eq10}
\mc{I}(t)\sim\begin{pmatrix}|\epsilon(t)|^{-\alpha}&|\epsilon(t)|^{b-1}\\|\epsilon(t)|^{b-1}&|\epsilon(t)|^{-\gamma}\end{pmatrix}
\end{equation}
where $\gamma=d \nu-2 b$, and $\alpha$ is the critical exponent corresponding to changes in temperature.

For the sake of simplicity we will now assume that only one intensive parameter, $\lambda(t)$, is varied. Thus, we can express the $(1\times 1)$-dimensional Fisher information matrix in terms of the general susceptibility $\mc{X}(t)$,
\begin{equation}
\label{eq11}
\mc{I}(t)=\mc{X}(t)=\mc{X}_0\,|\epsilon(t)|^{-\Lambda}
 \end{equation}
where $\Lambda$ is the critical exponent corresponding to the varied control parameter, e.g., for varied magnetic fields we have $\Lambda=\gamma$, and for processes with time-dependent temperatures $\Lambda=\alpha$.

The Kibble-Zurek hypothesis predicts that far from the critical point, $|t|\gg \hat{\tau}$, the dynamics is essentially adiabatic, and hence $\la W_\mrm{ex}\ra$ has non-vanishing contributions only in the impulse regime, $|t|\leq\hat{\tau}$, cf. Fig.~\ref{fig:plot_KZ}. Therefore, we can write,
\begin{equation}
\label{eq12}
\la W_\mrm{ex}\ra\simeq\lambda_c^2\,\int_{-n \hat{\tau}}^{n \hat{\tau}}dt\,|\dot{\epsilon}(t)|^2\,\tau(t)\,\mc{X}(t)
\end{equation}
where $\lambda(t)=\lambda_c (1-\epsilon(t))$ and $n>1$ is a small, real constant \footnote{We included the small, real constant $n>1$ to guarantee that no non-negligible contributions to the excess work are neglected.}. Employing Eqs.~\eqref{eq02} and \eqref{eq11} it is then a simple exercise to show that
\begin{equation}
\label{eq13}
\la W_\mrm{ex}\ra=\frac{2\lambda_c\,\mc{X}_0\,n^{-z\nu-\Lambda+1}}{z\nu+\Lambda-1}\,\tau_0^{\frac{2-\Lambda}{z\nu+1}}\,\tau_Q^{\frac{\Lambda-2}{z\nu+1}}\,.
\end{equation}
Equation~\eqref{eq13} constitutes our main result. We have shown that for systems that are driven at constant rate through a critical point the excess work, $\la W_\mrm{ex}\ra$, universally scales like,
\begin{equation}
\label{eq14}
\la W_\mrm{ex}\ra\sim\tau_Q^{\frac{\Lambda-2}{z\nu+1}}\,,
\end{equation}
which explicitly depends on the critical exponent $\Lambda$ corresponding to how the system is driven. This behavior is in full agreement with thermodynamics, since thermodynamic work is a \emph{process dependent} quantity \cite{Callen1985a}. In other words, Eq.~\eqref{eq14} expresses the fact that the excess work depends on \emph{how} the system is driven through the critical point, whereas the typical domain size $\hat{\xi}$ \eqref{eq05} is independent on the choice of the intensive control parameter.

\paragraph*{KZ-scaling in nonequilibrium systems}

The remainder of this analysis is dedicated to a slightly more general situation. We now consider any thermodynamic system whose dynamics is described by the Fokker-Planck equation,
\begin{equation}
\label{eq16}
\begin{split}
\pd_t\,p(x,t)&= \pd_x \left[-f(x) +D \left(x-m(T)\right)\right]\, p(x,t)\\
&+\pd_x \left[ T\, g(x) \,\pd_x g(x)\right]\, p(x,t)\,,
\end{split}
\end{equation}
where $f(x)$ is a conservative force  and $g(x)$ is a space-dependent diffusion coefficient. Equation~\eqref{eq05} is a mean-field description of an interacting lattice, where the interaction strength between two lattices sites is determined by $D$ \cite{VandenBroeck1994,Broeck1994}.  For such lattices the average position per lattice site $m(T)=\la x\ra$  is identical to the magnetization and hence constitutes the order parameter, which is determined self-consistently by \cite{Broeck1994}
\begin{equation}
\label{eq17}
\int dx\, p_\mrm{ss}(x, m) \,x=m(T)\,,
\end{equation} 
where $p_\mrm{ss}(x, m)$ \footnote{Note that generally $p_\mrm{ss}(x, m)$ is not a Boltzmann-Gibbs equilibrium distribution of the form $p_\mrm{eq}(x)\propto \e{-\beta V(x)}$ with $V(x)=-\int^x\td x' f(x')$. Therefore, even in the stationary state Eq.~\eqref{eq16} describes a true non-equilibrium system.} is the stationary solution of Eq.~\eqref{eq16}. 

Such systems are particularly interesting since for specific choices of $f(x)$ and $g(x)$ \cite{Broeck1994} they exhibit ``noise-induced'' phase transitions with $m(T<T_c)>0$ and $m(T>T_c)=0$ for a critical temperature $T_c$. Note that Eq.~\eqref{eq16} is an effective, mean-field description for discrete lattice models \cite{Broeck1994}. Hence, standard considerations of the KZM apply. 

In the following, we will be interested in purely temperature driven processes, and hence the total work vanishes, $\la W\ra=0$. In this case, the maximum available work theorem \eqref{app:eq03} becomes,
\begin{equation}
\label{app:eq05}
\la \Sigma\ra \equiv \Delta S-\beta \Delta E\geq 0\,.
\end{equation}
Thus, it will be convenient to continue the analysis in terms of the total entropy production $\la \Sigma\ra$.

 Note, that generally the irreversible entropy production, $\la\Sigma\ra$, is given by  \cite{Williams2008,Esposito2010d,Esposito2010,VanDenBroeck2010a,Deffner2012} 
\begin{equation}
\label{eq18}
\la \Sigma\ra=\Delta\mc{H}-\int dt\, \tr{\dot{p}(t) \,\beta(t) V}\, ,
\end{equation}
where $\Delta\mc{H}$ is the change of the Shannon information entropy with $\mc{H}(t)=-\tr{p(t) \lo{p(t)}}$. Here $\tr{\dots}=\int\td x \dots $ denotes an integral over configuration space and the dot denotes the derivative with respect to time. It is then easy to see that the total entropy production \eqref{eq06} can be separated into two terms \cite{Esposito2010d,Esposito2010,VanDenBroeck2010a},  $\la \Sigma\ra=\la \Sigma_\mrm{ad}\ra +\la \Sigma_\mrm{nad}\ra$ -- into the adiabatic entropy production
\begin{equation}
\label{eq19}
\la \Sigma_\mrm{ad}\ra =-\int dt\, \tr{\dot{p}(t)\, \lo{p_\mrm{ss}(t)/p_\mrm{eq}(t)}}
\end{equation}
and the nonadiabatic entropy production
\begin{equation}
\label{eq20}
\la \Sigma_\mrm{nad}\ra =-\int dt \,\tr{\dot{p}(t) \lo{p(t)/p_\mrm{ss}(t)}}
\end{equation}
where $p_\mrm{ss}(t)=p_\mrm{ss}(x,m(t))$ is the instantaneous stationary solution of Eq.~\eqref{eq16}, i.e., the stationary solution corresponding to the instantaneous value of $m(t)$.

In such nonequilibrium situations the part of the entropy production that corresponds to the excess work is $\la \Sigma_\mrm{nad}\ra$, which vanishes in the limit of quasistatic driving for which $p(t)\equiv p_\mrm{ss}(t)$ \cite{Esposito2010d,Esposito2010,VanDenBroeck2010a,Riechers2016}. In complete analogy to Eq.~\eqref{eq09} it can be shown that $\la \Sigma_\mrm{nad}\ra$ is given as a quadratic form of the Fisher information matrix \cite{Mandal2016,Bonanca2017}. Thus, we have
\begin{equation}
\label{eq21}
\la \Sigma_\mrm{nad}\ra \sim \tau_Q^{\frac{\Lambda-2}{1+z\nu}}\,,
\end{equation}
which follows from the same arguments as the ones leading to Eq.~\eqref{eq14}.

\paragraph*{Example: Noise induced phase transitions}

\begin{figure}
\includegraphics[width=.48\textwidth]{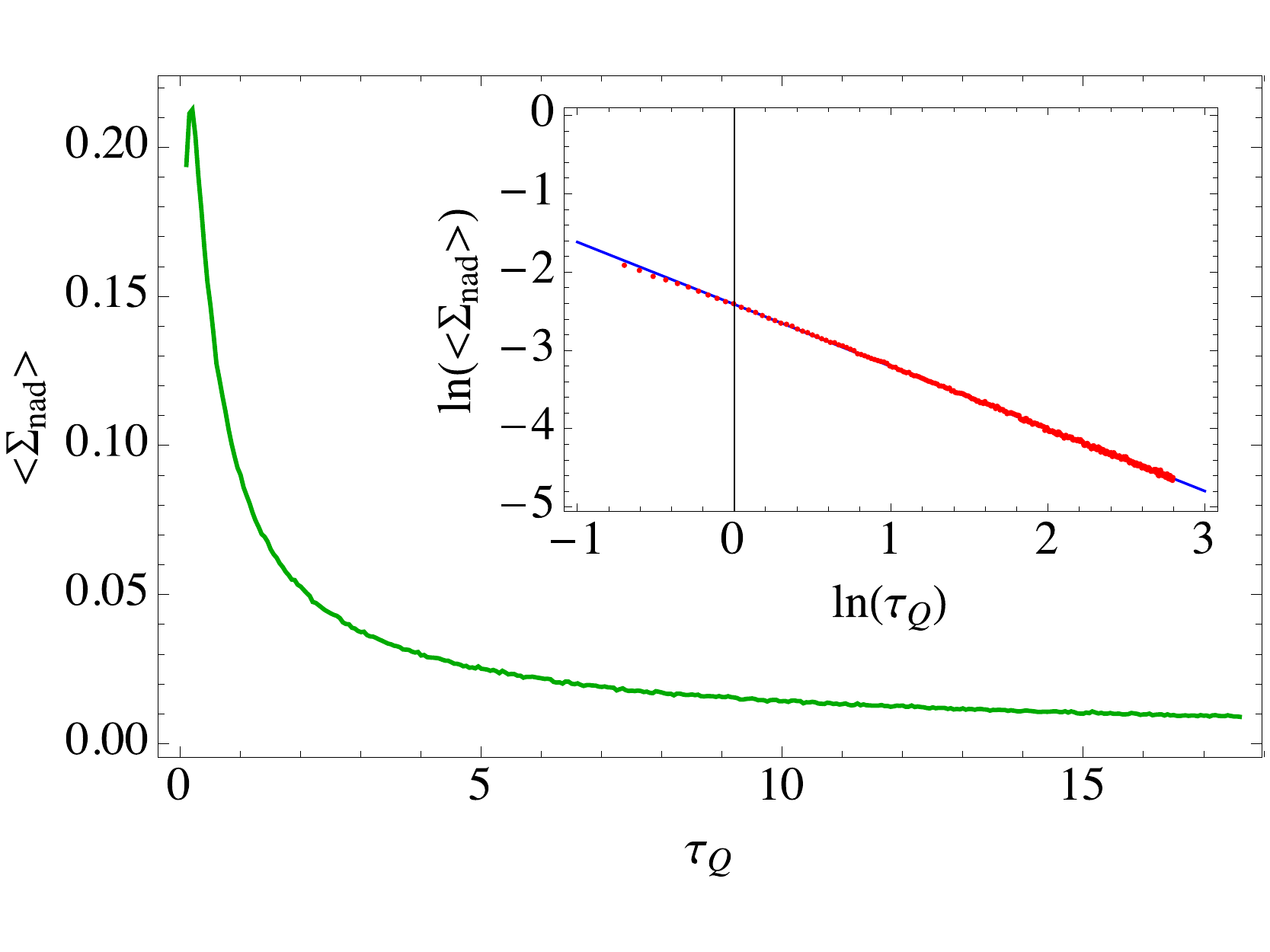}
\caption{\label{fig:entropy_1}(color online) Nonadiabatic entropy production \eqref{eq20} for Eq.~\eqref{eq22} with $D=15$, and for a quench from $T_i=8$ to $T_f=4$ (where we have $T_c\simeq 6.19$). The inset is a logarithmic plot (red dots) together with a linear fit (solid, blue line), from which we determine a scaling exponent of $0.78$. }
\end{figure}

The first system for which a noise-induced phase transition was found is given by \cite{Broeck1994},
\begin{equation}
\label{eq22}
f(x)=-x-2\,x^3-x^5 \quad \mathrm{and}\quad g(x)=1+x^2\,.
\end{equation}
We have solved the corresponding dynamics numerically for $D=15$, for which we found a phase transition at $T_c\simeq 6.19$. The resulting entropy production for a linear quench from $T_i=8$ to $T_f=4$ is plotted as a function of $\tau_Q$ in Fig.~\ref{fig:entropy_1}. We observe that for very fast quench times $\tau_Q\ll 1$ the nonadiabatic entropy production behaves irregularly, whereas for slower quenches $\la \Sigma_\mrm{nad}\ra$  exhibits clear polynomial behavior. The  behavior for short quenches can be understood by recalling the underlying assumptions of the KZM \cite{DelCampo2014}. The mechanism predicts that systems driven through a phase transition at finite rate experience first an adiabatic regime, before close to the critical point the dynamics becomes impulse-like. For quenches that are too fast the adiabatic regime cannot be established, and nonequilibrium excitations are created irregularly, i.e., not in accordance with the KZM -- the nonadiabatic entropy production behaves irregularly.

It has been shown that the noised-induced phase transitions for Eq.~\eqref{eq22}  belongs to the Kardar-Parisi-Zhang-universality class with $\nu=1$ and $z=3/2$ \cite{Toral2011}, and for varying the temperature we further have $\Lambda=0$ \cite{Broeck1994,VandenBroeck1994}. Thus, Eq.~\eqref{eq21} predicts,
\begin{equation}
\label{eq23}
\la \Sigma_\mrm{nad}\ra \sim \tau_Q^{-4/5}\,.
\end{equation}
The inset of Fig.~\ref{fig:entropy_1} shows a logarithmic plot together with a linear fit over three orders of magnitude. Numerically we find an exponent of $0.78$, which is in perfect agreement with the universal theory.

\begin{figure}
\includegraphics[width=.48\textwidth]{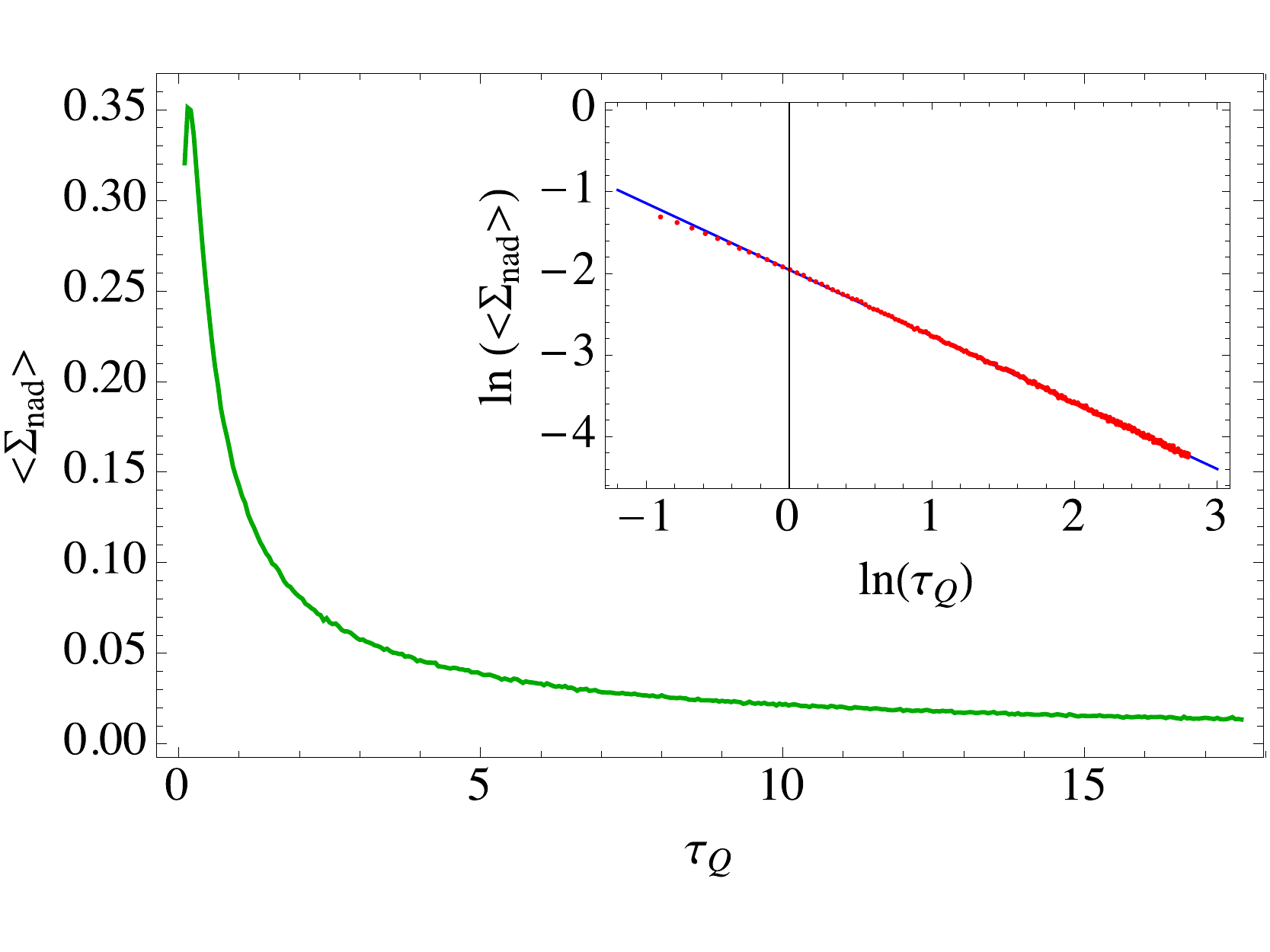}
\caption{\label{fig:entropy_2}(color online) Nonadiabatic entropy production \eqref{eq20} for Eq.~\eqref{eq24} with $D=15$, and for a quench from $T_i=11$ to $T_f=4.5$ (where we have $T_c\simeq 7.91$). The inset is a logarithmic plot (red dots) together with a linear fit (solid, blue line), from which we determine  scaling exponent of $0.81$. }
\end{figure}

As a second example we analyze
\begin{equation}
\label{eq24}
f(x)=-3/2\, x +x^3-x^5 \quad \mathrm{and}\quad g(x)=1+x^2\,.
\end{equation}
It has been found that Eq.~\eqref{eq24} induces much richer thermodynamic behavior \cite{Mueller1997}. In particular, there is a parameter range for which the model exhibits a first order phase transition, and a range where the phase transition is second order \cite{Mueller1997}. The second order phase transition is expected to also  belong to the KPZ-universality class \cite{Toral2011}.  The resulting nonadiabatic entropy production \eqref{eq20} is plotted in Fig.~\ref{fig:entropy_2} for a quench from $T_i=11$ to $T_f=4.5$. The inset shows again a logarithmic plot together with a linear fit over three orders of magnitude, from which we obtain the KZ-exponent $0.81$. Again, our numerical finding is in perfect agreement with the universal prediction \eqref{eq23}.

\paragraph*{Concluding remarks}

In the present analysis we have achieved two major results: (i) we have extended arguments of the KZM to quantify the universal scaling behavior of the excess work and irreversible entropy production; (ii) we have verified the universal theory and the KZM in noise-induced phase transitions. Thus our treatment generalizes the scope of the KZM to systems, for which notions such as domain walls or  topological defects, e.g., in the Bose-Hubbard model \cite{DelCampo2014,Gardas2017}, loose their clear meaning. On the conceptual side, the present work unifies the paradigms of two independently developed theories to describe nonequilibrium processes -- the Kibble-Zurek mechanism and Stochastic Thermodynamics.

\begin{acknowledgements}
It is a pleasure to thank Bart\l omiej Gardas for insightful discussions, and Wojciech H. Zurek for many years of mentorship and getting me interested in the Kibble-Zurek mechanism. SD acknowledges support by the U.S. National Science Foundation under Grant No. CHE-1648973.
\end{acknowledgements}

\bibliography{kz_work_bib}

\end{document}